\documentstyle[preprint,prd,aps,epsf]{revtex}

\newcommand{\postscript}[2]
{\setlength{\epsfxsize}{#2\hsize}
\centerline{\epsfbox{#1}}}

\tighten

\begin{document}

\draft

\title{ On Supersymmetric $b - \tau$ Unification, Gauge Unification,\\ and Fixed Points}

\author{Nir Polonsky}
\address{
Sektion Physik der
Universit$\ddot{\mbox{a}}$t M$\ddot{\mbox{u}}$nchen \\
Lehrstuhl Prof.\ Wess\\
Theresienstrasse 37\\
D-80333 M$\ddot{\mbox{u}}$nchen\\  Germany}

\maketitle

\begin{center}
{LMU-TPW-96-04}
\end{center}

\abstract{The equality assumption of the $b$ and $\tau$
 Yukawa couplings
at the  grand-unification scale can strongly 
constrain the allowed  parameter space
of supersymmetric models.
We examine the constraints
in the  case that there is a discrepancy 
$ \gtrsim 10\%$ in the gauge
coupling unification assumption
(which necessarily implies large perturbations at the grand scale).
The  constraints are 
shown to diminish in that case
[most significantly so if  $\alpha_{s}(M_{Z}) \approx 0.11$].
In particular, the requirement that the $t$ Yukawa coupling, $h_{t}$, 
is near  its quasi-fixed point may not be necessary.
We discuss the colored-triplet 
threshold as a simple example of a  source for the discrepancies,
and comment on its possible implications.
In addition, we point out that
supersymmetric (as well as  unification-scale)  
threshold  corrections to $h_{t}$ 
shift the fixed-point curve in the
$m_{t} - \tan\beta$ plane. 
The implications for the prediction of  the Higgs boson mass 
are briefly discussed.
}

\section{Introduction}
Unification of the $b$ and $\tau$ Yukawa couplings  \cite{yu} is known to be
consistent with the assumption
of low-energy supersymmetry \cite{susyyu}. 
However, the allowed parameter space depends sensitively 
on the exact value of the strong coupling 
$\alpha_{s}(M_{Z})  = 0.12 \pm 0.01$ 
used in the calculation \cite{bbo}.
In particular, using the results from gauge coupling unification to 
calculate the  $b$ and $\tau$ Yukawa couplings, 
$h_{b}$ and $h_{\tau}$, respectively,
strongly constrains the allowed range of the Higgs sector parameter
$\tan\beta \equiv \langle H_{2} \rangle / \langle H_{1} \rangle$
to $\tan\beta \sim 1$ or $\tan\beta \gg 1$ \cite{lp2,bcpw}.

Gauge coupling unification (including low-energy threshold corrections
but neglecting corrections at the grand-unification scale) 
generically implies $\alpha_{s}(M_{Z}) \gtrsim 0.13$  
and $\alpha_{s}(M_{G}) \sim 0.04$ \cite{lp4,bp} (where
$M_{G}$ denotes the unification point).
The one-loop\footnote{In our numerical calculations of gauge
and Yukawa couplings we will follow the procedure of Ref.\
\cite{lp2} using two-loop renormalization group equations
[three-loop equations for $\alpha_{s}(Q < M_{Z})$].
The procedure is extended in a straightforward manner
to include low-energy corrections to $m_{b}$ (see below).} 
expression for the weak-scale $b$ to $\tau$ mass ratio is
\begin{equation}
\frac{m_{b}(M_{Z})}{m_{\tau}(M_{Z})} \sim 
0.9\left[ \frac{\alpha_{s}(M_{Z})}{\alpha_{s}(M_{G})}\right]^{\frac{8}{9}} \times Y,
\label{mbmtau}
\end{equation} 
where the $0.9$ factor is from hypercharge renormalization, 
$Y < 1$ is a complicated
function of the Yukawa couplings, which is important for large couplings,
and $m_{\tau}(M_{Z}) = 1.75$ GeV.
Eq. (\ref{mbmtau}) and gauge unification imply (when neglecting $Y$) the prediction
$m_{b}(M_{Z}) \sim 4.5$ GeV. In comparison, the allowed 
(one standard deviation) range is  $m_{b}(M_{Z}) \lesssim 3.2$ GeV \cite{mb}
(but because of low-energy renormalization 
the upper bound is a function of $\alpha_{s}$).
The QCD corrections are thus too large 
and need to be compensated  by either large Yukawa coupling which diminish $Y$
(and also the prediction for $\alpha_{s}$) \cite{bbo,lp2,bcpw,largey} 
or finite one-loop supersymmetric threshold corrections to  $m_{b}$ 
(that are proportional to $\tan\beta$) \cite{hrs,wright}.
Both mechanisms can be realized  in the large $\tan\beta$ regime.
On the other hand,
in  the small $\tan\beta$ regime only the former is relevant,
and the allowed region  is strongly constrained in $\tan\beta$
by requiring for the top Yukawa coupling 
$h_{t}(m_{t}) \gtrsim 1.1$, i.e., that $h_{t}$ is near
its quasi-fixed point \cite{fp,qfp}. 
It is interesting to note that
for $\tan\beta \sim 1 $
the Higgs sector imitates that of the Standard Model (SM)
and contains a light  SM-like Higgs boson\footnote{
This is when considering finite QCD corrections (but see a discussion below)
to $m_{t}$ and re-summation of leading logarithms, 
which are the two most important
higher-order corrections. 
The formal one-loop bound does
not account for these effects by definition, and is higher by $10 - 15$ GeV
 (for example, see \cite{lp3}).
I thank Howard Haber for the discussion of this point. See also \cite{twoloop}.},
$m_{h^{0}}^{\mbox{\tiny one-loop}} \lesssim 100$ GeV,
which is within reach of LEPII \cite{hd,bbop,lp3}. 
Hence, in this minimal framework, 
Higgs boson searches contain information about Yukawa
unification.

However, the large predicted values of $\alpha_{s}(M_{Z})$ 
(note that the prediction increases quadratically
with  $m_{t}$) are somewhat uncomfortable phenomenologically \cite{shif}.
Particularly so, if the $Z \rightarrow b\bar{b}$ width is significantly larger
than what is predicted in the SM, as is currently implied by experiment \cite{lang}.
(In that case, the predicted 
$\alpha_{s}$ is typically subject to large and positive low-energy
threshold corrections \cite{lp4}, which further aggravate
the potential problem.)  
Low-energy corrections could have a large and negative contribution 
to the $\alpha_{s}$ prediction
only if $(a)$ the low-energy spectrum is extremely heavy and degenerate,
i.e., the correction parameters\footnote{
The leading logarithm correction to $\alpha_{i}^{-1}(M_{Z})$ is given by 
$(-\delta b_{i} /2\pi)\ln(M_{i}/M_{Z})$ where $\delta b_{i} = 25/10, 25/6, 4$ for 
$i = 1,\,2,\,3$, respectively. 
$\alpha_{1,\,2,\,3}$ denotes the hypercharge (normalized
by $5/3$), weak and strong couplings, respectively.}
 $M_{1}, M_{2}$ and $M_{3}$ defined in Ref.\ \cite{lp1}
are large and equal, or 
$(b)$  $M_{2} \gg M_{1},\, M_{3}$  (see Figure  5a of Ref.\ \cite{lp1}).
The former mechanism is not very likely, as it implies
a degeneracy between colored ($M_{3}$) and 
non-colored (e.g., $M_{2}$) particles,
contradictory  to the different nature of the 
radiative corrections in both sectors\footnote{
When including the radiative corrections, the 
leading-logarithm correction to the prediction is 
typically proportional to the 
supersymmetric Higgs mass $\mu$ \cite{bcpw} and is more
likely to be positive. It is negative if
$\mu$ is very large. 
On the other hand, a large $\mu$ typically implies large mixing between
left- and right-handed scalars and possibly a light scalar.
The inclusion of finite  corrections 
results now in a positive shift of the one-loop correction \cite{lp4,bp}.
Because of this anti-correlation
between the finite and logarithmic corrections,
 it is very difficult to obtain a negative 
one-loop correction \cite{elp}.
The Roszkowski-Shifman
proposal described below 
does not affect the proportionality to $\mu$, but only
its coefficient \cite{bcpw}.}.
It was suggested, however, that the latter
mechanism could be realized if the QCD 
gauge fermions (the gluinos) are much lighter than 
the weak gauge fermions (the winos) \cite{sr}. 
While possible, this would imply that supersymmetry
breaking is transmitted to the observable sector 
at a much lower scale than the breaking
of the grand-unified group:
If the supersymmetry breaking is transmitted to the 
visible sector gravitationally at
Planckian scales, then the ratio of the different 
gaugino masses is dictated by
the grand-unified symmetry to be approximately equal to that of the 
respective gauge couplings\footnote{If the gauge 
kinetic function is grossly non-minimal,
then this relation, and also gauge coupling unification,
can be altered \cite{ed}.}.
Such models \cite{dnns} must contain new 
exotic matter beyond the minimal supersymmetric
extension (MSSM), and are not discussed in this work
(but see Ref.\ \cite{cm}).

Thus, if indeed $\alpha_{s}(M_{Z})  \lesssim 0.12$, 
then  one expects (aside from the above mentioned caveat)
significant perturbations to the naive
grand-unification relations at the unification scale.
This is a crucial point when discussing Yukawa unification.
It is straightforward to show that low-energy corrections
to the $\alpha_{s}$ prediction constitute only a second-order perturbation
in the $m_{b}(M_{Z})$ prediction \cite{lp2}
(but they could affect the $M_{Z} - m_{b}$ renormalization). 
However, corrections at the unification scale
are multiplied by a large logarithm and 
can, depending on the way in which they propagate into 
the $m_{b}/m_{\tau}$ relation, correct the $m_{b}$ prediction significantly.

In this note we investigate the possible implications of
such a scenario to Yukawa unification.
Our purpose is not to define the allowed parameter space
with any high precision, but rather examine whether
such a precision is possible beyond the minimal 
framework (which is not favored by the data).
In Section II,  we discuss two examples of corrections:
nonrenormalizable operators (NRO's) and  colored triplet thresholds.
(We also include in our numerical analysis low-energy corrections to 
$m_{b}$.) We examine the allowed parameter space as a function of
$\alpha_{s}$ and of $h_{t}$. The latter is a useful measure of the parameter
space which is independent of the size of the low-energy corrections
to $m_{t}$, discussed in Section III. 
We find that the gap between the allowed small and
large $\tan\beta$ regions  is a sensitive function of $\alpha_{s}$,
the low-energy corrections to $m_{b}$ (and thus, the soft parameters),
$m_{t}$, and of the unification-scale perturbation to $h_{b}/h_{\tau}$.
Outside the minimal framework 
(which constrains $\alpha_{s}$ and the perturbations),
none of these parameters
is significantly constrained and the range of the 
allowed $\tan\beta \gg 1$ region is ambiguous.
In particular, the gap nearly vanishes if $\alpha_{s}(M_{Z}) \sim 0.11$,
or if the unification scale perturbation is ${\cal{O}}(20\%)$.
Even though one can, in general,  distinguish two different branches,
the distinction is less significant as the gap diminishes, undermining
the motivation to consider one branch rather than the other.
Thus, the strong constraints on $b-\tau$ unification are intimately
linked to the large values of $\alpha_{s}$ predicted in the minimal framework.
In Section III we discuss 
the sensitivity of the $h_{t}$ fixed-point
curve to different threshold and other corrections,
and stress that one-loop supersymmetric corrections
to $h_{t}$ are as important as the standard QCD correction.
We conclude in Section IV, where we also 
point out the
implications to the prediction of the
Higgs boson mass in Yukawa unified models.

\newcommand{\vs}{\mbox{\large \it vs.} }
\section{Gauge \vs Yukawa unification}

Before discussing  examples of possible unification-scale corrections
to the $\alpha_{s}$ prediction, it is important to realize
the smallness of typical couplings at that scale and its implications:
\begin{itemize}
\item $\alpha_{s}(M_{G}) \sim 0.04$. 
Because of the QCD enhancement\footnote{
This is similar to the scaling between the QCD and weak scales
that drastically reduces large uncertainties 
in the $\alpha_{s}$ measurements at ${\cal{O}}(1$ GeV$)$  when
propagated to $M_{Z}$.
(The smaller coupling is compensated in our case by a larger logarithm.)} 
of small unification scale perturbations 
in the value of $\alpha_{s}(M_{G})$, the allowed $\sim \pm 8\%$
range of $\alpha_{s}(M_{Z}) = 0.12 \pm 0.01$ corresponds to only a 
$\sim \pm 3\%$
(or $\sim \pm 0.0015$) 
range at the unification scale.
\item
$h_{\tau}(M_{G}) \sim 1/100\cos\beta$
($y_{\tau} = h_{\tau}^{2}/4\pi \sim 10^{-5}$), and 
similarly $h_{b}(M_{G}) \sim 0.01 $ (for $\tan\beta \sim 1$).  
\newline 
In extrapolating $h_{\tau}$ we used the near flatness of its renormalization
curve (for not too large $\tan\beta$) . Note also that when using the data
as boundary conditions, $h_{b}(M_{G}) < h_{\tau}(M_{G})$ 
by ${\cal{O}}(10^{-3})$. 
In Fig.\ 1 it is shown that typically [for $\alpha_{s}(M_{Z}) = 0.12$]
$(h_{b} - h_{\tau})/h_{\tau} \sim -0.2$ at $M_{G}$.
The ratio is $\sim -0.3$ for $\alpha_{s}(M_{Z}) = 0.13$
and $\sim -0.1$ for $\alpha_{s}(M_{Z}) = 0.11$.
\end{itemize}
Hence, a small numerical perturbation 
constitutes a large percentile perturbation.

The smallness and near flatness of 
$h_{\tau}$ is of particular importance in our case \cite{ck}.
It implies that small shifts in $h_{\tau}(M_{G})$
correspond to an apparent large violation of 
$h_{b} - h_{\tau}$ unification. 
One can visualize this as shifting the initial point of a nearly
flat line (the $h_{\tau}$ renormalization curve). 
A small shift  can drastically change its  intersection 
with the moderately sloped $h_{b}$ renormalization curve
(the slope of the QCD renormalized $h_{b}$ curve decreases at high energies
where the couplings are small),
leading to an apparent
(or effective) unification point which could be 
many orders of magnitude below $M_{G}$.
(Recall that the renormalization 
curve is a function of $\ln{Q}$ and not $Q$.)
One can control such shifts by requiring that the apparent 
Yukawa-unification scale is not more than two or three orders
of magnitude below $M_{G}$ \cite{lp2}.
Such a constraint, however, is not motivated 
if one allows large shifts elsewhere
[e.g., in $\alpha_{s}(M_{G})$].
If one eliminates such (``no-conspiracy'') constraints,
then there could be corrections of ${\cal{O}}(100\%)$
in the case that $h_{b}$ and  $h_{\tau}$
are still numerically small (i.e., for $\tan\beta \sim 1$). 
On the other hand, from Fig.\ 1,  one observes that already 
${\cal{O}}(20\%)$ corrections remove many of the constraints.
We return to this point below.
 
Next, we elaborate on possible corrections to $\alpha_{s}$.
One mechanism that could possibly shift 
$\alpha_{s}(M_{G})$ is gravitational smearing
(i.e., gravitationally induced NRO's), originally 
proposed as a non-perturbative mechanism
\cite{nro1,ed} and later realized as an efficient 
perturbation (or smearing) to
unification relations \cite{nro2,nath,nro3}.
Requiring that the effect is perturbative typically 
constrains the 
coefficient of the (leading) operator
such that the absolute value of the correction 
to the $\alpha_{s}(M_{Z})$ prediction (which depends on the correlated shifts
of all three gauge couplings)
is  $\lesssim 0.010-0.015$.
(The exact number depends on the group theory structure.)
One could argue for a larger correction, depending
on the perturbativity criterion imposed.
On the other hand, one typically expects a smaller correction,
e.g., in Ref.\ \cite{lp1} it was estimated that the absolute
value of the correction is $\lesssim 0.006$.
The correction can be propagated to the
$m_{b}/m_{\tau}$ ratio as a constant shift in $\alpha_{s}(M_{Z})$ \cite{lp2}
(see also Ref.\ \cite{nath}).
In addition, other operators could now shift 
the boundary conditions of other couplings, e.g., $h_{\tau}(M_{G})$,
generating the perturbations discussed above.

A different mechanism for lowering  the $\alpha_{s}$ prediction   
is by introducing an SU(5) breaking
between (colored and non-colored) heavy 
chiral supermultiplet thresholds.
In extended models many candidates could exist 
(for examples, see Refs.\ \cite{lp1,mpm,mar,moh,flip,raby}).
However, the most obvious candidate is the colored triplet Higgs, $T$,
that has to be split from the light Higgs doublets
(see also Ref.\ \cite{k}). 
Indeed, the doublet-triplet splitting
problem, even though solvable by fine-tuning
of the superpotential and of the scalar potential,
calls for non-generic solutions 
that may affect the properties of the triplet threshold \cite{split}.
We consider this generic threshold as an example only.

Typically, one assumes  $M_{T} \gtrsim M_{G}$ 
so that the loop-level  (dimension-five)
colored-Higgsino mediated proton decay \cite{proton1}
is sufficiently suppressed \cite{proton2}.
Nevertheless, the effectiveness  of $M_{T} \sim 10^{-2}M_{G}$
in lowering the prediction for $\alpha_{s}$ may suggest
a different mechanism for suppression 
of the dimension-five proton decay operator.
One possibility\footnote{Other possibilities involve suppression due
to symmetries \cite{proton3}, group-theory \cite{mpm,proton4,proton5,hc}, and the 
structure of the soft terms \cite{fpt}.}
is that all Yukawa couplings of $T$ 
are suppressed \cite{dvali}, in which case the only correction to
Yukawa unification is via
the modification of $\alpha_{s}$\footnote{
Ignoring proton decay constraints, one could entertain the idea
that an intermediate scale triplet drives
$\alpha_{s}(M_{Z}) < 0.11$, which is then corrected to
$\alpha_{s} (M_{Z})> 0.11$ by low-energy thresholds.}$^{,}$\footnote{
In the light triplet models of Ref.\ \cite{dvali} the correction is proportional
to the logarithm of the triplet to (new) doublet mass ratio.},
\begin{equation}
\Delta_{\alpha_{s}} \sim \frac{9\alpha_{s}^{2}(M_{Z})}{14\pi}\ln\frac{M_{T}}{M_{G}}.
\label{delalphas}
\end{equation} 
A different possibility is that some of the Yukawa couplings
of $T$ to the third generation are  not suppressed.
This assumption is particularly motivated here, since naive Yukawa unification
is successful
only in the case of the third family, and thus, provides
no information on the Yukawa couplings (and mixing angles)
of the two light families.
If these are the only couplings which are not suppressed, then proton 
decay  constraints on  $M_{T}$ are diminished.
In addition to diminishing $\alpha_{s}(M_{T})$,
the triplet threshold in this case 
$(i)$ introduces a Yukawa coupling
correction to $h_{b}/h_{\tau}$, 
$(ii)$ shifts the $h_{t}$ fixed point (see Section III),
and $(iii)$ renormalizes the soft parameters
(i.e., the scalar potential)
corresponding to the third family,
an effect
which is particularly important for the 
mass of the scalar $\tau$,  which could become too
light or tachionic \cite{talk}.
From $(i)$ one has a 
$\sim \{1 - [(h^{2}_{t}(M_{G})/(16\pi^{2})]\ln(M_{T}/M_{G})\}$
correction factor to (\ref{mbmtau}) \cite{wright}, which can be absorbed
as a shift in the boundary conditions.
(We will include it explicitly
in the numerical integration, i.e., in the numerical calculation of $Y$.)
From $(iii)$, 
there could be an enhancement\footnote{In  principle,  
one could obtain a (model-dependent) lower bound
on $M_{T}$, independent  of proton decay and of the 
$\alpha_{s}$ prediction.}
of low-energy  lepton-number violation processes\footnote{
We find \cite{up}, for example, an enhancement as large as two
orders of magnitude  (for $M_{T}/M_{G} \gtrsim 10^{-3}$)
to the $\mu \rightarrow e \gamma$
branching ratios of the models considered in Ref. \cite{bhs}.}, e.g.,
$\mu \rightarrow e \gamma$ \cite{talk,up}.

In fact, both mechanisms, the operators
and the triplet threshold, may be  linked.
Perturbations of some form or another are required 
in order to explain the failure of Yukawa unification 
for the two light families.
One common mechanism to generate 
 these perturbations is NRO's
which are either gravitational or higher-symmetry 
remnants. Such operators most probably 
shift also
the third family Yukawa couplings, and could
allow only extra 
suppressed couplings for the colored triplet.

We examine the parameter space in Figs.\ 2--3, where we fixed
$m_{t}^{pole} = 170$ GeV 
(consistent  with direct \cite{top} and indirect \cite{lang}
determinations).
In order to examine the smearing of the allowed\footnote{
We require $4.00  \leq m_{b}(m_{b}) \leq 4.45$ GeV
(e.g., see Ref.\ \cite{mb}). 
More points would be allowed had we imposed this constraint,
but for  $m_{b}(m_{b}^{pole} \sim 5$ GeV) rather
than for  $m_{b}(m_{b})$.
For a discussion, see also Refs.\ \cite{lp2,hrs}.} 
$\tan\beta$ range for $\alpha_{s}(M_{Z}) = 0.12$,
we require in Fig.\  2 that $b-\tau$ unification at the $\alpha_{1} - \alpha_{2}$
unification point ($M_{G} \approx 3 \times 10^{16}$ GeV) holds
to a precision of either $5\%,\,15\%$ or $25\%$.
In practice, this would typically mean $h_{b}(M_{G}) \rightarrow
 0.8h_{\tau}(M_{G})$, leading to a better agreement
with the data. 
For example, a perturbation of  $15\%$ 
[or $h_{b}(M_{G}) \sim 0.85h_{\tau}(M_{G})$]
corresponds in some cases
 to an apparent Yukawa-unification point as low as  $10^{10}$ GeV.
Low-energy corrections to $m_{b}$ \cite{hrs} are also included
and  calculated explicitly assuming, for simplicity,
``universal'' boundary conditions
to the soft parameters at the grand scale\footnote{
For simplicity, we do not include renormalization
effects above $M_{G}$ \cite{pp}.}, 
and radiative symmetry breaking,
agreement with experimental lower bounds on the masses (and
an imposed upper bound of $\sim 2$ TeV), and
using a monte-carlo scan of the parameter space
(for further details, see \cite{lp3}).
We account for NRO's (or other corrections
whose main effect is to shift  $\alpha_{s}$ at high energies)
by fixing $\alpha_{s}(M_{Z}) = 0.120 $
[and $\alpha_{s}(M_{Z}) = 0.110,\, 0.130$ in Fig.\ 3].
For comparison, we also show the respective 
allowed points when  not
including the low-energy corrections (diamonds).
 
As implied by Fig.\ 1, for a  $25\%$ perturbation,
no constraints exist on small $\tan\beta$.
It is interesting to note that it is extremely difficult  to find 
very large $\tan\beta$ solutions.
The exclusion of $\tan\beta \gtrsim 45 \sim m_{t}/m_{b}$
results from  the simultaneous requirement
of radiative symmetry breaking and 
acceptable threshold
corrections to $m_{b}$ (and may be overcome by excessive tuning
of parameters \cite{copw,rs}, in particular, 
in  non-universal schemes \cite{cw,rs}).
When  not including the low-energy corrections (diamonds),
these points are again allowed, 
but the intermediate $\tan\beta$ range is excluded [unless there is
a $ \gtrsim {\cal{O}}(20\%)$ perturbation].
The extreme tuning (for small perturbations)
of very small and very large
$\tan\beta$ solutions (e.g., see diamonds in Fig.\ 2) 
may suggest that the allowed region
of intermediate $\tan\beta$ solutions is  preferred. 
However, one has to be cautious, as such solutions depend
sensitively on the soft parameters\footnote{
There is also a correlation (which we do not treat in this work)
between the $m_{b}$ correction and the size of the chargino loop
contribution to $b \rightarrow s\gamma$, and a negative correction
typically implies an enhancement of the $b\rightarrow s\gamma$ rate \cite{copw}.
This effect is generally important for $\tan\beta \gtrsim 25-30$
and a too high $b \rightarrow s\gamma$ rate may exclude
some of the allowed points in that region, depending on
the charged Higgs mass.}.
In Fig.\ 4 we show the possible low-energy corrections to $m_{b}$,
where points which constitute the $5\%$ perturbation curve in Fig.\ 2
are indicated by bullets. Only a small fraction of points has
the required $\sim -20\%$ correction.
Therefore, for small perturbations, all solutions for Yukawa
unification require some tuning.
(In principle, one could distinguish three allowed regions,
but because of their complimentary nature, we will keep
identifying both the intermediate and the very large $\tan\beta$
branches as the large $\tan\beta$ solution.)
 
We further examine solutions with small ($5\%$) perturbations in Fig.\ 3,
where we fix $\alpha_{s}(M_{Z}) = 0.110,\, 0.130$.
The latter is roughly the value one  would get when requiring 
gauge coupling unification and $M_{T} \gtrsim M_{G}$, i.e.,
the minimal framework.   
We also present curves 
requiring gauge coupling unification but fixing  
$M_{T} = 10^{15},\, 10^{14}$ GeV [$\alpha_{s}(M_{Z}) \approx 0.118, 0.112$, 
respectively].
(The triplet threshold is included numerically and the correlation between
the shifts in $\alpha_{s}(M_{Z})$, $\alpha_{s}(M_{G})$
and $M_{G}$ \cite{lp2} is automatically accounted for.)

In the minimal framework, even when including the low-energy corrections,
the two branches, $\tan\beta \sim 1.3$ and $\tan\beta \gtrsim 15$
are clearly distinguished.
However, the small $\tan\beta$ solution is
extremely tuned in this case
because of the large QCD correction (see Section III)
and because of  the $M_{Z} - m_{b}$ QCD renormalization.
[A ${\cal{O}}(1-2\%) $ low-energy corrections can now exclude
an otherwise consistent solution.]
While a significant gap remains in this case,
it is smeared almost completely for $\alpha_{s} \sim 0.110$.
It is worth stressing, however,
that some gap remains (for small perturbations) in all cases.
Thus, one can still distinguish
two allowed branches, as in the minimal framework.
This is because of the fixed point relative insensitivity for 
corrections to  $\alpha_{s}$ and 
the proportionality of the $m_{b}$ corrections to $\tan\beta$,
which lead to only negligible smearing of the 
$\tan\beta \sim 1$ branch.
Nevertheless, smearing of the large $\tan\beta$ branch
down to $\tan\beta \sim 8 (4)$ for $\alpha_{s}(M_{Z}) \sim 0.120 (0.110)$
significantly diminishes the excluded region, as well as
undermines arguments (based on Yukawa unification)
in favor of the $\tan\beta \sim 1$ branch.
Furthermore, as $m_{t}^{pole}$ increases, the $h_{t}$ fixed-point
curve is flatter in $\tan\beta$, further diminishing the gap (see Fig.\ 5).
Also, given the smallness of $h_{b}$ and $h_{\tau}$ for
$\tan\beta \sim 1$, 
${\cal{O}}(20\%)$ perturbations are reasonable,
as discussed above, and  the $\tan\beta \sim 1$ branch  could also be 
smeared (see Fig.\ 2).

In Fig.\ 5 we allow $m_{t}^{pole} = 180 \pm 12$ GeV \cite{top} (with a Gaussian 
distribution)  and show the allowed values of the top Yukawa coupling
$h_{t}$ as a function of $\tan\beta$ for $\alpha_{s}(M_{Z}) = 0.120$
and a $5\%$ perturbation.
(Note that for large values of $m_{t}^{pole} \gtrsim 190 - 200$ GeV,
$h_{t}$ could be near its fixed point for intermediate
values of $\tan\beta$.) The requirement $h_{t} \gtrsim 1.1$ holds
for $\tan\beta \lesssim 8$.
This is a reflection of the respective excluded region (the gap)
in Fig.\ 2 where $m_{t}^{pole} = 170$ GeV  
(and  $h_{t} \lesssim1.1$ for  $\tan\beta \gtrsim 1.4$).
The fact that now there is no gap 
is due to the higher values
of $m_{t}^{pole}$.

\section{The  fixed-point curve}

Points near the $h_{t}$ fixed-point were shown
above to provide a solution to $b$-$\tau$ Yukawa unification.
That solution is the least sensitive to either enhancement
or suppression of the low-energy corrections to $m_{b}$
(the sensitivity grows with $\alpha_{s}$, as discussed above). 
However,  the solution
is a result of  the large numerical
value of $h_{t}$ only, and because of the $h_{t}$
convergence to its fixed-point value
this result is relatively insensitive to $\alpha_{s}(M_{Z})
= 0.12 \pm 0.01$.
The translation of this value to a curve in the 
$m_{t}^{pole} -\tan\beta$ 
plane contains a few ambiguities,
which are worth recalling.

In fact, this is only a quasi-fixed point  \cite{qfp} (convergence from above).
If the low-energy $h_{t}$ exceeds its fixed point value,
then it becomes non-perturbative at some higher scale.
In a consistent calculation
the quasi fixed-point has to be defined numerically, e.g., that 
renormalization from two-loops is smaller than a certain
fraction of that from one-loop. This leads, e.g., to the condition 
$h_{t} \lesssim 3$
at all scales below the cutoff scale \cite{bbo}.   
Therefore, the cutoff scale for the calculation
enters the definition.  For example, using $10^{18} $ GeV
rather than $M_{G}$ as a cutoff, leads [in SU(5)]
to the requirement
$h_{t}(M_{G}) \lesssim 2$, shifting the fixed point curve
to slightly higher values of $\tan\beta$. 
(In fact, there may be another quasi-fixed point $h_{t} \sim 2$
at $M_{G}$ \cite{bhs}.)
In addition, the fixed-point value
of $h_{t}$ depends on the other large couplings in the
renormalization group equations, i.e., $\alpha_{s}$.
The lower $\alpha_{s}$ is, the lower is that value, and again,
the curve slides to slightly larger values  of $\tan\beta$
(e.g., this can be seen in  Fig.\  3).

If there are other large couplings,
i.e., new large Yukawa couplings
(or a large number of new couplings), then the fixed-point
value of $h_{t}$ also changes.
The quasi-fixed point is reached by a cancellation of
gauge and Yukawa terms.
Since the size of the former is roughly fixed, 
any new Yukawa coupling modifies the upper bound on
all other Yukawa couplings (that enter the same set 
of renormalization group equations).
New Yukawa couplings could renormalize
$(i)$ $h_{\tau}$, $(ii)$ $h_{b}$, and $(iii)$ $h_{t}$.
In most examples all three are relevant and 
a fixed-point value of $h_{t} < 1$  is possible
(i.e., $\tan\beta$ slides to larger values) while
still maintaining Yukawa unification.
Some examples include
$(a)$ low-energy singlets \cite{lp3,king}, $(b)$ fourth family \cite{four1,four2},
and $(c)$ baryon and lepton number violating couplings \cite{blv}.

A most interesting case is that
of  $(d)$ an intermediate-scale  right-handed neutrino 
where only $(i)$ and $(iii)$
occur.  Before its  decoupling at the intermediate-scale,
the new Yukawa coupling, $h_{\nu}$,  renormalizes $h_{\tau}$
in the same way 
that $h_{t}$ renormalizes $h_{b}$.
The two Yukawa corrections roughly cancel in the ratio
[assuming $h_{\nu}(M_{G}) \approx h_{t}(M_{G})$],
and the Yukawa correction function $Y$ in (\ref{mbmtau})
is closer to unity (depending on the right-handed
neutrino scale), unless $h_{b}$, itself, is significantly large  \cite{nur,moh}.
The small $\tan\beta$ solution is excluded in this case,
regardless of the exact location of the $h_{t}$ fixed point.

The generic heavy threshold corrections 
follow similar patterns.
The adjoint field, like the singlet [case $(a)$],
 is coupled to the ``Higgs-leg'' of the
Yukawa operators, and the effect cancels in the
$h_{b}/h_{\tau}$ ratio \cite{lp2}. 
However, it also affects $h_{t}$, and hence, affects
$h_{b}/h_{\tau}$ indirectly. However,  unlike the low-energy singlet case,
the indirect correction here is suppressed by a small logarithm.
It could  shift the fixed-point 
if its coupling to the Higgs doublets, which renormalizes $h_{t}$,
is large [i.e., in SU(5) it is the case that the color triplet is heavy],
and its self coupling (that determines its own mass)
is small.
The color triplet has lepto-quark
couplings that unify with $h_{t}$,
and is a special  example of $(c)$.
Because of its large mass 
(i.e., the small logarithm) the effect is again moderate. 
We find that for $M_{T} \gtrsim 10^{14}$ GeV
the fixed-point value
of $\tan\beta$ increases 
(including the modification of the $\alpha_{s}$ prediction)
by less than 0.18 
(and less than 0.06 for a fixed value of  $\alpha_{s}$). 
 
Lastly, supersymmetric threshold corrections to $m_{b}$ play a
crucial rule in expanding the allowed parameter space:
They generate the allowed intermediate $\tan\beta$ region
in the case of small perturbations.
Similar corrections have been shown to affect
other parameters \cite{pr}, an observation
which is related to renewed interest \cite{rh,hc}
in (weak-scale) radiative fermion masses \cite{radfm}.   
In fact, it is doubtful that one can consider
predictions for the SM fermionic sector parameters independently
from the supersymmetric spectrum parameters.
The corrections that are relevant for our discussion are those for
the $m_{t}^{pole}/m_{t}^{\overline{\mbox{ \tiny DR}}}$
ratio
($\overline{\mbox{DR}}$ stands for the dimensional-reduction scheme), 
\begin{equation}
h_{t} = \frac{
m_{t}^{{\mbox{\tiny $\overline{\mbox{ \tiny DR}}$}}}}{174
{\mbox { GeV}}} \frac{\sqrt{1 + \tan^{2}\beta}}{\tan\beta}.
\label{ht} 
\end{equation}
We defined the parameter 
$m_{t}^{\overline{\mbox{ \tiny DR}}}$
to absorb all threshold corrections, i.e., at one-loop
\begin{equation}
m_{t}^{\overline{\mbox{ \tiny DR}}} =
m_{t}^{pole}\left[ 1 - \Delta^{t}_{\mbox{\tiny QCD}}
- \Delta^{t}_{\mbox{\tiny SUSY-QCD}} - \Delta^{t}_{\mbox{\tiny EW}}
\right],
\label{mt}
\end{equation}
where\footnote{ 
One also needs to include a $\Delta^{b}_{\mbox{\tiny QCD}} = 
[{1}/{3}][{\alpha_{s}(M_{Z})}/{\pi}]$ when converting $m_{b}(M_{Z})$
from its $\overline{\mbox{DR}}$ definition to its
modified minimal-subtraction definition, which is the relevant one
for $m_{b}(m_{b})$.
This correction is important, e.g., for $\alpha_{s}(M_{Z}) = 0.13$.}
\cite{mv}
\begin{equation}
\Delta^{t}_{\mbox{\tiny QCD}} = \frac{5}{3}\frac{\alpha_{s}(m_{t})}{\pi},
\label{mtqcd}
\end{equation}
and $\Delta^{t}_{\mbox{\tiny EW}}$ includes electroweak
and Yukawa contributions \cite{wright,mtdr} that we neglect hereafter.
$\Delta^{t}_{\mbox{\tiny SUSY-QCD}}$ 
includes new QCD contributions in the MSSM (which are
only implicitly dependent on $\tan\beta$),
that have been calculated using three- \cite{wright} and two-point \cite{mtdr} 
functions and shown
to be potentially of the order of magnitude of
$\Delta^{t}_{\mbox{\tiny QCD}}$.  
Recently, it has been further shown \cite{fpt}
that $\Delta^{t}_{\mbox{\tiny SUSY-QCD}}$ does not 
have  a fixed sign\footnote{The leading logarithm terms
agree in sign with $\Delta^{t}_{\mbox{\tiny QCD}}$,
but the overall sign is model dependent.}
and introduces
a significant ambiguity in the fixed-point curve.
In particular, this correction can be more
important than the $\sim 2\%$ two-loop QCD contribution
to (\ref{mtqcd}) that many authors include while neglecting
supersymmetric loops.

In Fig.\ 6 we examine the corrections for the point
($m_{t}^{pole}$, $\tan\beta$) = (170 GeV, 1.4), i.e.,
in the vicinity of the ``naive'' fixed-point curve, 
and for $\alpha_{s}(M_{Z})  = 0.12$
(using the vertex formalism
of Ref.\ \cite{wright} and imposing the same assumptions
on the parameter space as above).  
By fixing $h_{t}$ to its fixed-point value,
the corrections are absorbed in the invariant
combination $m_{t}^{pole}/\sin\beta$.
(Note that the corrections, though represented by
a mass parameter, are in fact corrections
to the Yukawa coupling.)
It is straightforward to absorb the corrections
in  $m_{t}^{pole}$ (vertical line), 
in which case  the correction in
our example
is $-2\% \lesssim \Delta^{t}_{\mbox{\tiny SUSY-QCD}}
\lesssim 5\%$ or  between $-3$ to 8 GeV. (The asymmetry is due to the fixed
sign of the leading logarithms). 
However, 
if $m_{t}^{pole}$ is known with high-precision, than
the corrections are to be absorbed\footnote{This is a similar procedure to
absorbing radiative corrections in the weak angle rather than
in $M_{Z}$.} 
in $\sin\beta$
(horizontal line). [A similar procedure could be used
to treat the uncertainty in $\alpha_{s}$ in (\ref{mtqcd}).]
The two-lines define a region in the parameter space
that corresponds to one point on the
``naive'' fixed-point curve.
Fig.\  5 is insensitive to this ambiguity, but
the interpretation of Figs.\ 2--3  is sensitive.
The ambiguity in $m_{t}^{pole}/\sin\beta$
diminishes the required tuning of the $\tan\beta \sim 1$
solutions (at the price of dependence on the the soft term) 
in a similar way to the smearing of the
large $\tan\beta$ solutions due to the corrections to $m_{b}$.
The correction (absorbed in $m_{t}^{pole}$) is 
shown in Fig.\ 7 for any $\tan\beta$
for $\alpha_{s}(M_{Z}) = 0.12$ (and requiring
$b -\tau$ unification with a $5\%$ perturbation).
The dependence on $\tan\beta$ 
is from the supersymmetric Higgs
mass $\mu = \mu(\tan\beta,\, ...)$, the left-right
$t$-scalar mixing, and a correlation between
the $m_{t}$ and $m_{b}$ corrections 
(which we do note explore in detail in this work).

\section{conclusions}

To conclude, the increasing value of the $\alpha_{s}$ prediction
significantly constrains
the allowed parameter space for Yukawa unification.
Yet, if $\alpha_{s}$ is significantly lower than predicted,
there exists a  significant perturbation at the grand scale, 
examples of which we discussed
in Section II. 
Such a perturbation creates an ambiguity which removes
many of the constraints on Yukawa unification.
The constraints were shown to be a sensitive function
of $\alpha_{s}$, unification-scale perturbations, 
and low-energy corrections to $m_{b}$  (and of $m_{t}$),
and nearly vanish for $\alpha_{s}(M_{Z}) = 0.11$ 
or  a ${\cal{O}}(20\%)$ low or high-scale correction.
From our figures one can obtain a qualitative description of 
the excluded region (the gap) in terms of the lower bound on
the large $\tan\beta$ branch (for $m_{t}^{pole}  = 170$ GeV), 
\begin{equation}
\tan\beta \gtrsim \frac{1}{2}\left\{ \frac{\left[\alpha_{s}(M_{Z}) - 0.100\right]}{0.001}
+ \frac{h_{b}(M_{G}) -  h_{\tau}(M_{G})}{0.010\times h_{\tau}(M_{G})}  \right\}.
\label{gap}
\end{equation}
(We assume that the left-hand side of (\ref{gap}) is $\geq 1$, 
otherwise $\tan\beta \geq 1$.)
Thus, the success of ``simple'' gauge unification 
[$\alpha_{s}(M_{Z}) > 0.12$]
and the constraints on Yukawa unification
are intimately linked, and the difference
between the predicted and measured $\alpha_{s}$ values can be viewed
as a sensitive measure of the typical size of 
perturbation at the unification scale.
We also pointed out the ambiguity in the location of
the fixed point and demonstrated the need to consider threshold corrections
to $m_{t}$  when discussing the fixed-point curve.
This also affects the $h_{t}$-perturbativity lower bound
on $\tan\beta$.

The required properties of the unification-scale perturbations,
which we simply assumed when discussing examples,
can, on the one hand, put severe constraints on model building
and enhance the predictive power in the high-scale
theory (see, for example, Ref.\ \cite{raby}).
On the other hand, it implies loss of some predictive
power in the low-energy theory, i.e., 
unlike the minimal framework,
now there are no generic predictions
but only  model-dependent ones
(which depend on additional parameters).
The loss of low-energy predictive power may be 
compensated in some cases
by the effects of threshold corrections 
(due to these perturbations) in the soft parameters
on flavor changing neutral current processes,
but these are again strongly model dependent.

Regarding the light Higgs boson mass $m_{h^{0}}$,
its lightness is due to the accidental proximity of
the $h_{t}$ fixed-point curve to the flat direction
in the Higgs scalar potenial for $\tan\beta = 1$.
The latter implies
$m_{h^{0}}^{\mbox{\tiny Tree}} < M_{Z}|\cos 2\beta|
\rightarrow 0$ near the fixed point curve.
If the curve slides to larger values of $\tan\beta$,
 $m_{h^{0}}^{\mbox{\tiny Tree}}$ increases.
However, unless new Yukawa couplings are introduced
[e.g., examples $(a) - (c)$ above],  the increase is
$\lesssim 10$ GeV, and since the mass $m_{h^{0}}^{\mbox{\tiny one-loop}}$
is a sum in quadrature of tree and loop terms, 
it has no significant ambiguity.
The ambiguity due to
$\Delta^{t}_{\mbox{\tiny SUSY-QCD}}$
is more of an interpretational ambiguity,
since $h_{t}$ (or $m_{t}^{\overline{\mbox{ \tiny DR}}} $)
is the relevant parameter for the calculation
of the loop correction in 
$m_{h^{0}}^{\mbox{\tiny one-loop}}$.
(As commented above, this is actually one of the more important
higher-order  refinements of the calculation.)
The prediction of 
$m_{h^{0}}^{\mbox{\tiny one-loop}}$
is thus insensitive to the corrections (if absorbed
in $m_{t}^{pole}$). 
However, the correspondence between 
$m_{t}^{pole}$ and $m_{h^{0}}^{\mbox{\tiny one-loop}}$
is now ambiguous.
We thus conclude  that, indeed, 
Higgs searches can probe the MSSM fixed-point region.
However, while this region may be motivated by
various reasons (not the least, the existence of a fixed-point
structure)\footnote{For example, 
it was recently suggested
that the only possibility to reconcile the $Z \rightarrow b\bar{b}$
discrepancy, mentioned above, with supersymmetric extensions
is if $\tan\beta \sim 1$ \cite{wk}. See also \cite{fpt}.},
the diminished gap between 
the two allowed branches
for Yukawa unification
undermines the uniqueness of the $\tan\beta \sim 1$ branch and the 
motivation to consider this region based on $b-\tau$ unification,
unless $\alpha_{s}$ is large and unification-scale perturbations are small.

\acknowledgments
This work was supported by a fellowship from the
Deutsche Forschungsgemeinschaft.
I thank Paul Langacker for his comments on the manuscript, 
and Zbigniew Pluciennik for comparison of some of his results for 
the $h_{t}$ corrections \cite{cpp2}.

\begin{figure}[h]
\label{fig:fig1}
\postscript{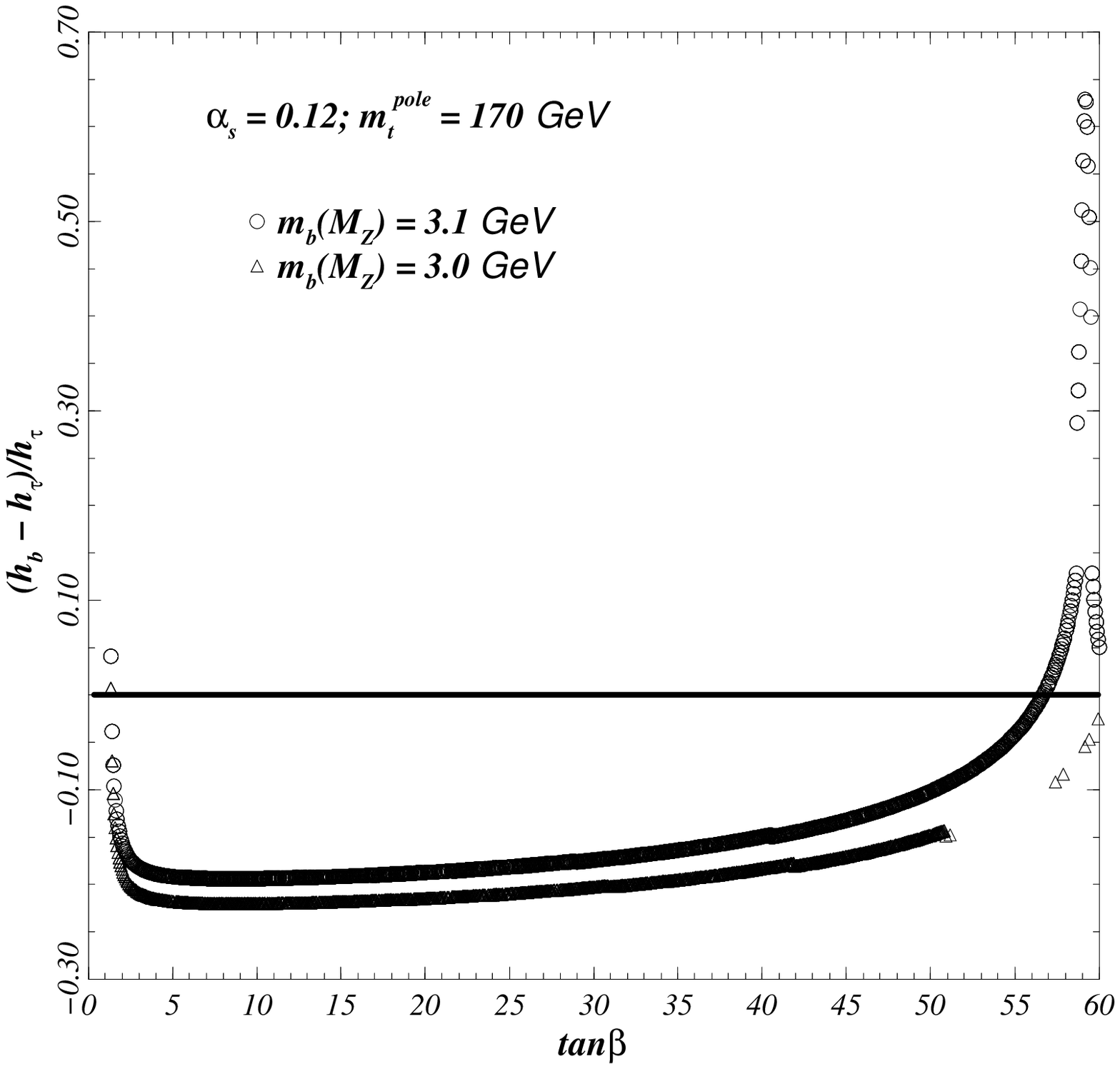}{0.9}
\caption{
The unification-scale difference
$h_{b} - h_{\tau}$ is shown in $h_{\tau}$ units
for $m_{b}(M_{Z}) = 3 $ GeV, $\alpha_{s}(M_{Z}) = 0.12$, 
$m_{t}^{pole} = 170$ GeV and as a function of $\tan\beta$.
For comparison, we also show the difference for
$m_{b}(M_{Z}) = 3.1 $ GeV (which for $\alpha_{s}(M_{Z}) = 0.12$
is inconsistent with $m_{b}(m_{b}) < 4.45 $ GeV).
Note the rapid change near the (naive) small and large 
$\tan\beta$ solutions, which is a measure
of the required tuning.}
\end{figure}

\begin{figure}[h]
\label{fig:fig2}
\postscript{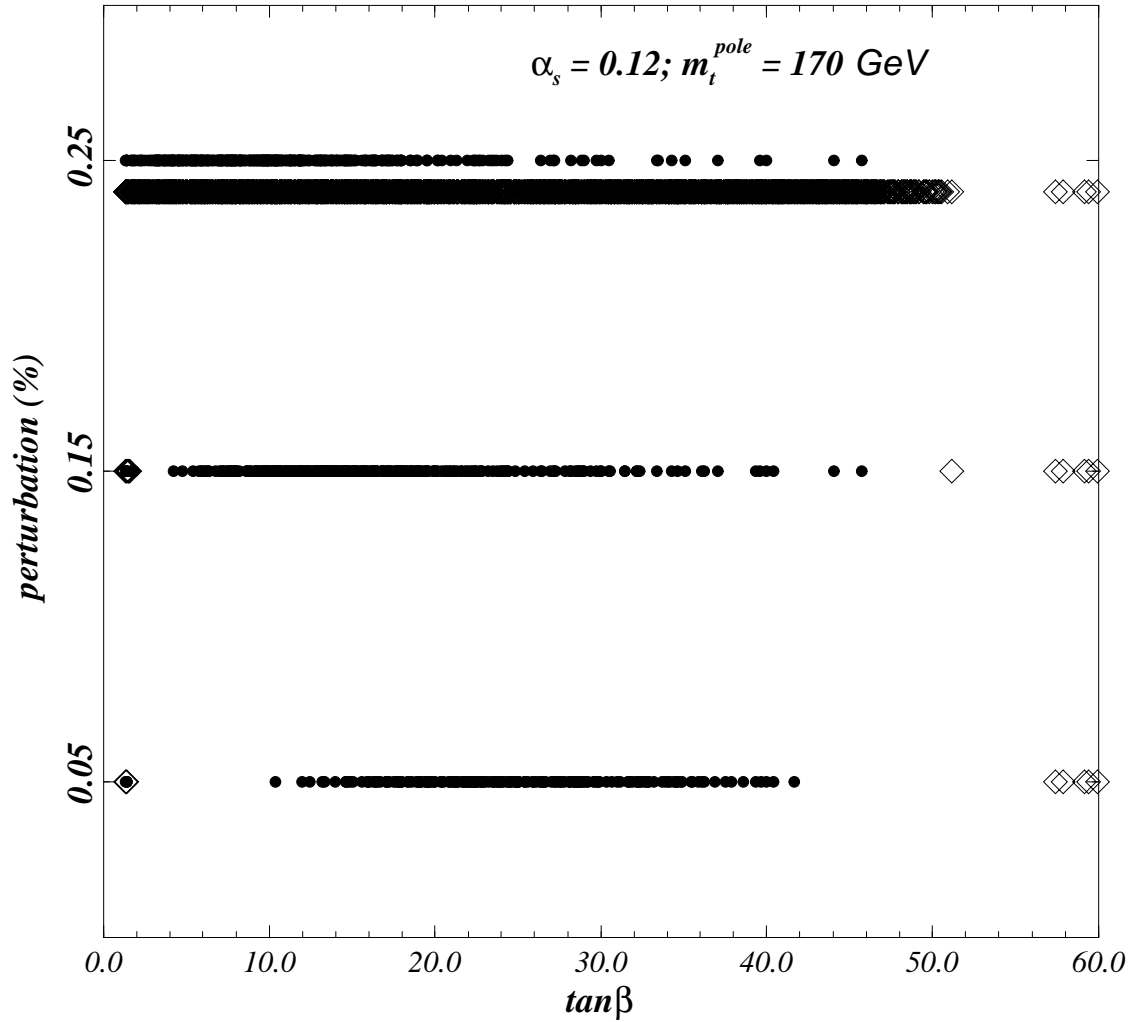}{0.9}
\caption{The MSSM points which are consistent with
$b-\tau$ unification for $\alpha_{s}(M_{Z}) = 0.12$ and
$m_{t}^{pole} = 170$ GeV
are shown as a function of $\tan\beta$
when including (bullets) and when omitting  (diamonds)
low-energy corrections to $m_{b}$.
The different curves correspond to
$h_{b}/h_{\tau} = 1 \pm 0.05, \, 1 \pm 0.15, \, 1\pm 0.25$,
at the unification scale, respectively.
(The two upper curves correspond to $1\pm 0.25$.)
}
\end{figure}

\begin{figure}[h]
\label{fig:fig3}
\postscript{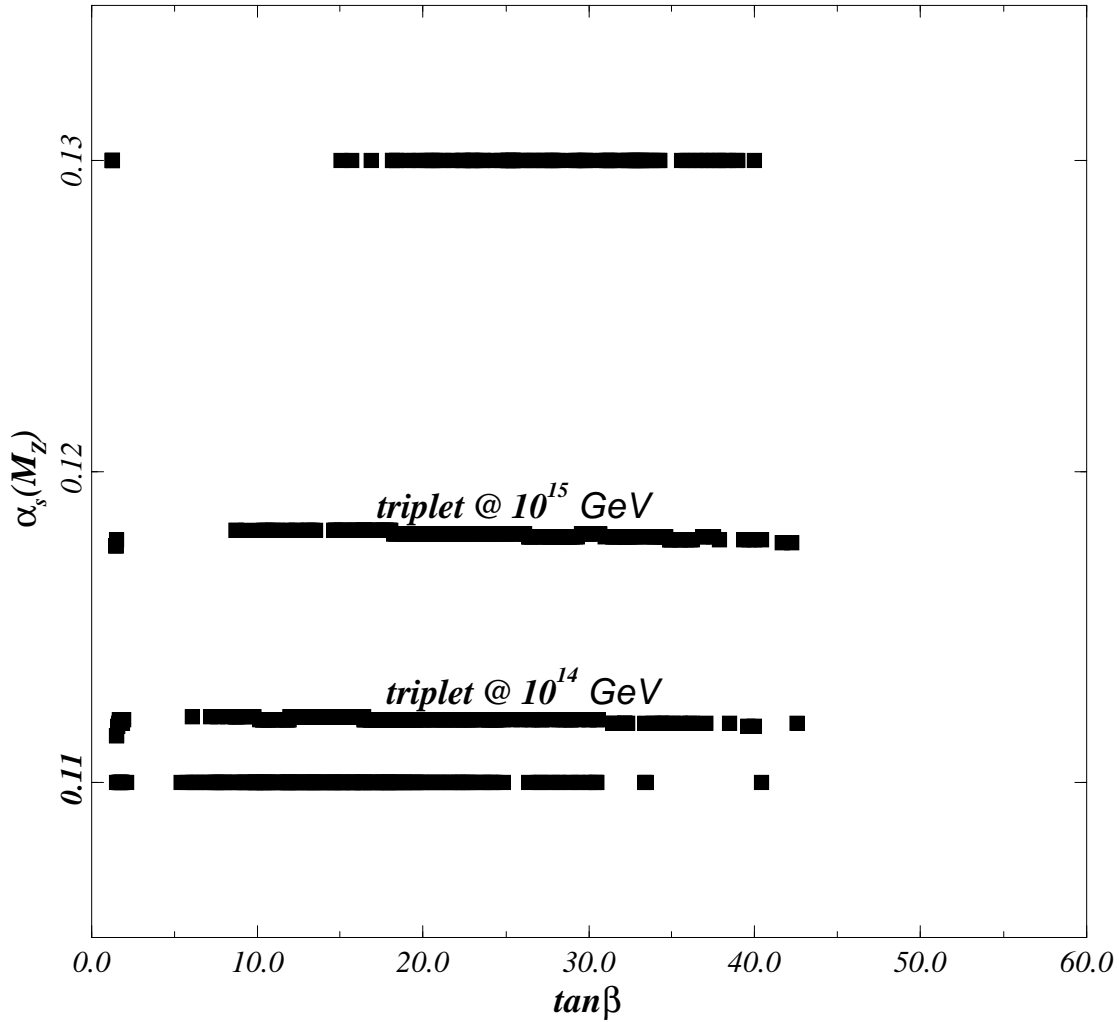}{0.9}
\caption{The MSSM points which are consistent with
$b-\tau$ unification for 
$m_{t}^{pole} = 170$ GeV
and when requiring 
$h_{b}/h_{\tau} = 1 \pm 0.05$
are shown as a function of $\tan\beta$
(including low-energy corrections to $m_{b}$).
The upper and lower curves correspond to 
$\alpha_{s}(M_{Z}) = 0.13,\,0.11$,  respectively.
In the two middle curves $\alpha_{s}(M_{Z}) $
is predicted when a colored triplet threshold
at $M_{T} = 10^{15},\,10^{14}$
GeV (with Yukawa couplings to the third family) 
is assumed (and accounted for in the numerical integration).
}
\end{figure}

\begin{figure}[h]
\label{fig:fig4}
\postscript{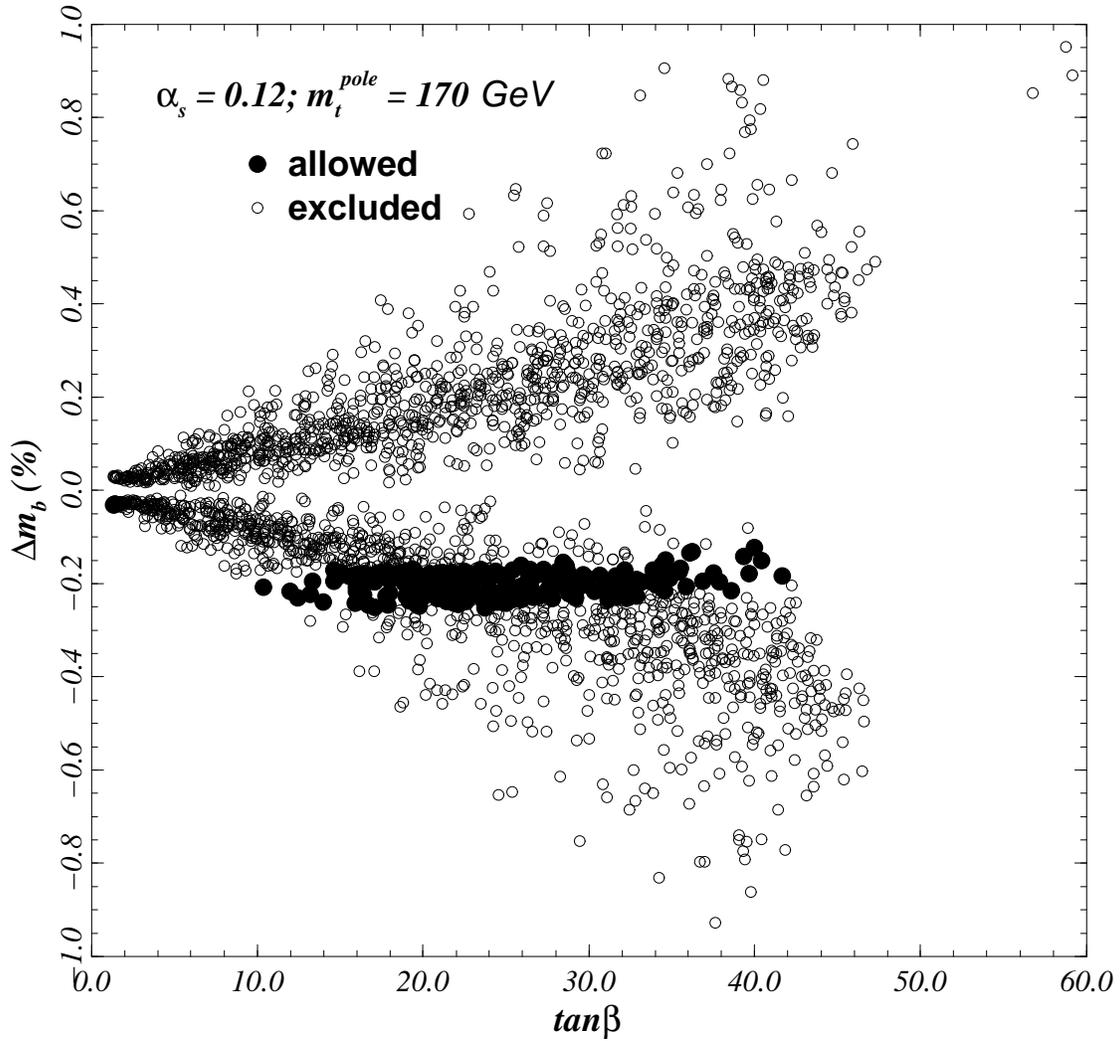}{0.9}
\caption{The low-energy threshold corrections 
to $m_{b}$ for the MSSM points considered in Fig. 2.
Only the points indicated by bullets correspond to
the $1\pm 0.05$ curve in Fig. 2.}
\end{figure}

\begin{figure}[h]
\label{fig:fig5}
\postscript{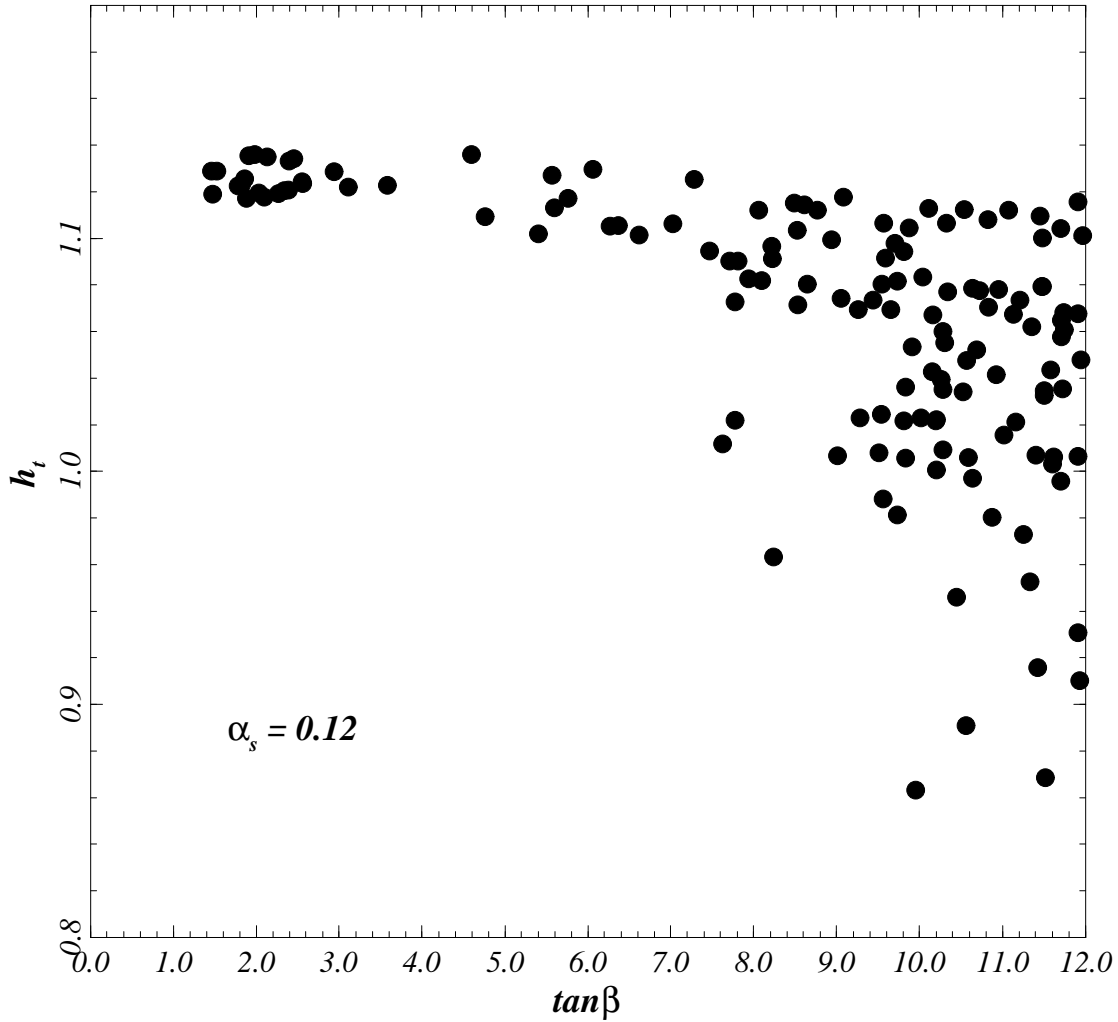}{0.9}
\caption{The MSSM points which are consistent with
$b-\tau$ unification for 
$\alpha_{s}(M_{Z}) = 0.12$,
$m_{t}^{pole} = 180 \pm 12$ GeV
and when requiring 
$h_{b}/h_{\tau} = 1 \pm 0.05$,
are shown as a function of $\tan\beta$ and of
the Yukawa coupling $h_{t}$
(which is calculated including only its QCD correction).
Because of the larger values of 
$m_{t}^{pole}$, larger values of $h_{t}$ (and thus, solutions
for $b-\tau$ unification)
are obtained for $\tan\beta > 2$.
The correspondence between $h_{t}$ and 
$m_{t}^{pole}$ could change when including the SUSY-QCD corrections
of section III.
}
\end{figure}

\begin{figure}[h]
\label{fig:fig6}
\postscript{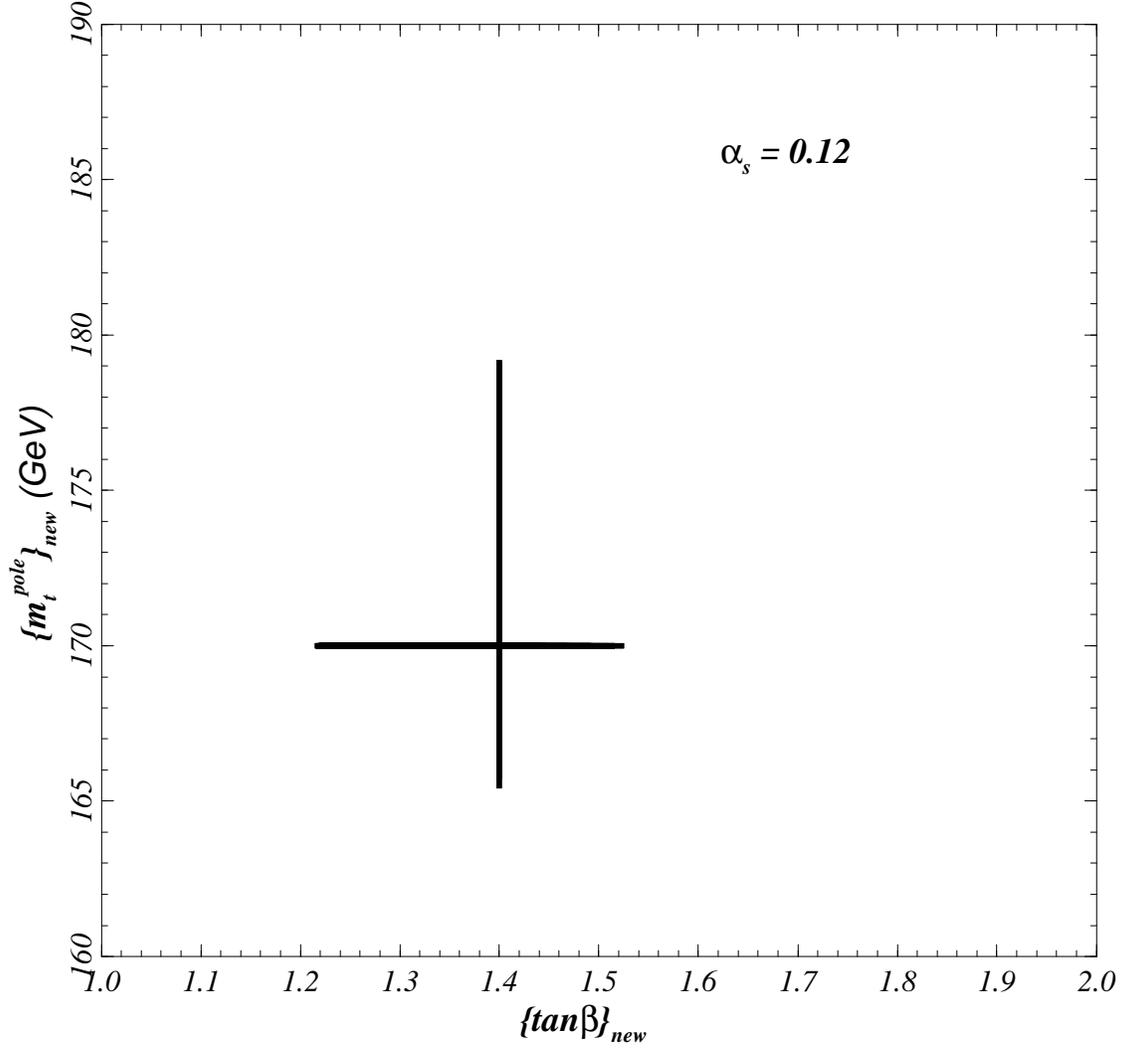}{0.9}
\caption{
The SUSY-QCD corrections to $h_{t}$
are absorbed in $m_{t}^{pole}$ (vertical line)
and  in $\tan\beta$ (horizontal line), smearing the naive point
$m_{t}^{pole} = 170$ GeV
and $\tan\beta = 1.4$
(assuming a fixed $h_{t}$ value and $\alpha_{s}(M_{Z}) = 0.12$).
$b-\tau$ unification is not required.
}
\end{figure}

\begin{figure}[h]
\label{fig:fig7}
\postscript{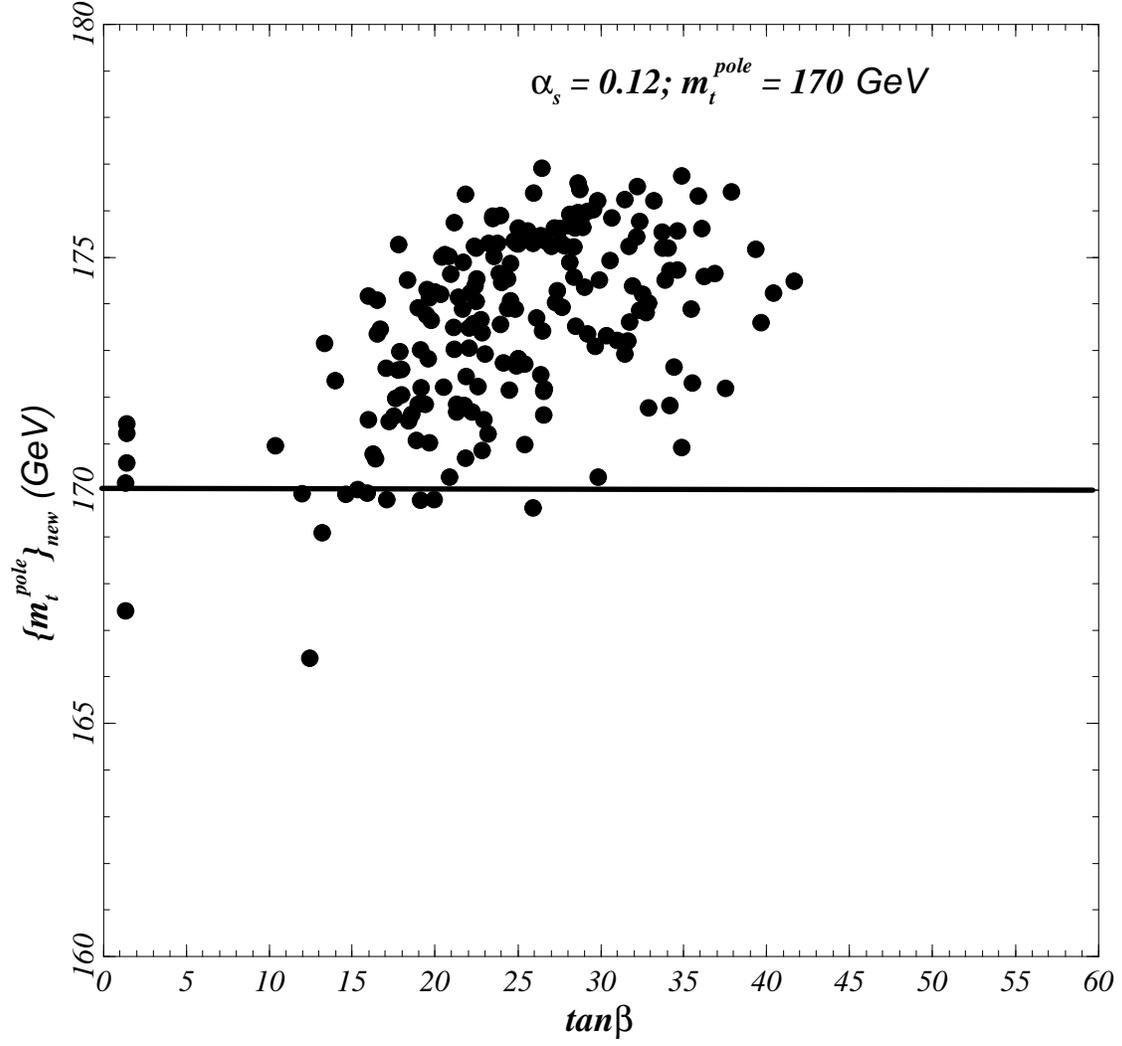}{0.9}
\caption{
The SUSY-QCD corrections to $h_{t}$
are absorbed in $m_{t}^{pole}$ 
for the points of the $1\pm 0.05$ curve in Fig. 2.}
\end{figure}

\end{document}